\documentclass[aps,prd,twocolumn,showpacs,groupedaddress]{revtex4}
\usepackage[dvips]{graphicx}
\usepackage{bm}
\usepackage{times}
\usepackage{amssymb}
\usepackage{amsmath}

\begin{document}
\bibliographystyle{myprsty}

\title{Laser--dressed vacuum polarization in a Coulomb field}

\author{A. I. Milstein  }
\author{I. S. Terekhov}
\affiliation{Budker Institute of Nuclear Physics, 630090 Novosibirsk, Russia}
\author{U. D. Jentschura }
\author{C. H. Keitel}
\affiliation{Max-Planck-Institut f\"ur Kernphysik, 
Saupfercheckweg 1, 69117 Heidelberg, Germany}

\date{\today}

\begin{abstract}
We investigate quantum electrodynamic
effects under the influence of an external,
time-dependent electromagnetic field, which mediates dynamic modifications
of the radiative corrections. Specifically, we consider 
the quantum electrodynamic vacuum-polarization tensor under the
influence of two external background fields: a strong laser 
field and a nuclear Coulomb field.
We calculate the charge and current densities induced by a 
nuclear Coulomb field in the 
presence of a laser field. We find the corresponding induced
scalar and vector potentials.
The induced potential, in first-order perturbation 
theory, leads to a correction to atomic energy levels.
The external laser field breaks
the rotational symmetry of the system.
Consequently, the induced charge density is not spherically symmetric,
and the energy correction therefore leads to a ``polarized Lamb shift.''
In particular, the laser generates an additional 
potential with a quadrupole moment. 
The corresponding laser-dressed vacuum-polarization 
potential behaves like $1/r^3$ at large 
distances, unlike the Uehling potential that vanishes 
exponentially for large $r$. The energy corrections are of 
the same order-of-magnitude for hydrogenic levels,
irrespective of the angular momentum quantum number.
The induced current leads to a transition dipole 
moment which oscillates at the second harmonic
of the laser frequency and is mediated by second-order harmonic 
generation in the vacuum-polarization loop.
In the far field, at distances $r \gg 1/\omega$ from the 
nucleus ($\omega$ is the laser frequency), the laser induces 
mutually perpendicular electric and magnetic fields, which give 
rise to an energy flux that corresponds to photon fusion leading to 
the generation of real photons, again at the second harmonic
of the laser. Our investigation 
might be useful for other situations where quantum field theoretic
phenomena are subjected to external fields of a 
rather involved structure.
\end{abstract}


\maketitle


\section{INTRODUCTION}
\label{intro}

This article is concerned with the calculation of
the vacuum-polarization tensor in the presence of two 
nontrivial background fields: a nuclear Coulomb field and a (possibly strong)
laser field. Particular emphasis is laid on the induced charge 
density and potential, as well as the induced current 
and the vector potential, and the resulting induced 
electromagnetic fields. As an application, we consider the 
energy shifts of atomic states in hydrogenlike systems,
and induced transition probabilities,
under the influence of the 
laser-dressed vacuum polarization. However, we would like to 
emphasize that the two latter aspects constitute merely 
example applications of the more general expressions derived in 
this article.

Before we indulge into this endeavour, let us briefly 
recall a few very well known facts about vacuum polarization.
It is well known that the Coulomb law becomes invalid at
distances comparable to the electron Compton wavelength
$\lambdabar_{\rm e} = 3.8 \times 10^{-11}\,{\rm cm}$.
This phenomenon is due to the excitation of virtual
electron-positron pairs from the vacuum, which polarize the
vacuum in a way approximately equivalent to a dielectric
medium. Indeed, at large distance, we see only a ``screened''
charge, which is less than the ``bare'' charge visible to
an observer who approaches the particle in question,
in our case mostly an atomic nucleus, to a very close
distance. The total charge density induced in the vacuum
has to be zero, and this also defines the renormalization
condition for the charge which can be defined via the low-energy
limit of the Thomson scattering cross section~(see~\cite{ItZu1980},
Chap.~7), as it affects two distant charge particles.
However, as we enter the cloud of the electron-positron
pairs that surround the charged nucleus, the effective charge becomes
larger, because one has to add to the ``renormalized'', physical
charge of the nucleus (seen at large distance),
a contribution due to the electron-positron
pairs that surround the nucleus at short distance.


So, the virtual excitations of the electron-positron
(or muon-antimuon) pairs naturally generate a ``fifth-force''
like Yukawa potential contribution to the Coulomb law,
which is however exponentially suppressed at distances
larger than the Compton wavelength, and by a quantum
electrodynamic (QED) perturbation theory parameter $\alpha$.

What happens to the vacuum-polarization effects in a laser field?
Evidently, the laser field, which is composed of many photons,
will modify the electron-positron loop as the laser photons
interact with the virtual quanta during their short lifetime
that is only of the order of $\lambdabar_{\rm e}/c =
1.3 \times 10^{-21}\,{\rm s}$. We find that in a
remarkably strong laser field, the modification of the
vacuum-polarization contribution to the Coulomb law can be
significant. In particular, the laser induces a rotational symmetry
breaking in the correction to the Coulomb law, which
leads to a polarized Lamb shift whereby states with different
magnetic quantum numbers receive different energy shifts.
The induced charge density, in a laser field, has a monopole and
a quadrupole component. The quadrupole part of the 
potential is not exponentially suppressed at large distances, but
tails off like a ``normal'' quadrupole potential, i.e.~as $1/r^3$. 

In part, our investigation follows ideas outlined previously in
the context of the analysis of the vacuum-polarization
tensor in a laser field~\cite{BaMiSt1975,BeMi1975}. Here, our focus
is on applications including a (possibly strong)
binding Coulomb field in addition to the laser field.
Another fundamental QED process, the self-energy under the
influence of an external laser field, for a free particle,
has previously been studied in~\cite{BaKaMiSt1975,BeMi1976}, and for a strongly
driven transition between bound states,
in~\cite{JeEvHaKe2003,JeKe2004aop,EvJeKe2004,JeEvKe2005}.
A more general context of our study might be defined as
``quantum field theory under the influence of external fields.''

From a naive point of view, one might expect that in a 
strong laser field, the charge density induced by the 
vacuum polarization loses the rotational symmetry, and that
an alignment along the laser propagation and polarization axes occurs.
However, the exact functional form of the 
induced laser-dressed contribution to the charge 
density [see Eq.~(\ref{phir}) below] is not accessible by symmetry
arguments. 
It is therefore indispensable to 
use a fully laser-dressed formalism,
where the quantum electrodynamic effects are incorporated into 
the propagators right from the start of the calculation.
As will be discussed below, the result of a fully
laser-dressed calculation confirms the intuitive picture,
according to which rotational invariance is broken 
in both linear as well as circularly polarized laser
fields, and a few unexpected asymptotic structures 
are found for the laser-induced fields.

This article is organized as follows: in Sec.~\ref{density},
we analyze the additional charge density induced by the external 
laser field (this potential is an addition to the 
``normal'' Uehling potential). The way from the charge density to the 
induced potential is detailed in Sec.~\ref{calcdensity}.
The asymptotics of the induced potential, in various regions 
from the atomic nucleus, are derived in Sec.~\ref{asymppotential}.
As an application, the energy shift of hydrogenic bound states
is considered in Sec.~\ref{energypotential}.

%
%
\section{INDUCED CHARGE DENSITY AND POTENTIAL}
\label{density}

%
%
\subsection{From the charge density to the potential}
\label{calcdensity}

The amplitude $\cal T$ of photon scattering
in the laser field has the form 
\begin{equation}
{\cal T}=\varepsilon_2^{\mu *}\,T_{\mu\,\nu}(k_2,k_1)\,\varepsilon_1^\nu\,.
\end{equation}
Here, the four-vectors $k_1$ and $k_2$ characterize the 
incoming and outgoing photon momenta, as they flow into
and out of the laser-dressed loop.
The polarization vectors of the photons
are $\varepsilon_1^{\mu}$ and $\varepsilon_2^\mu$.
The quantity $T_{\mu\,\nu}(k_2,k_1)$ is 
canonically identified as the vacuum-polarization tensor.
The calculation of the tensor $T_{\mu\,\nu}(k_2,k_1)$
is a rather involved calculation, but complete results,
valid for arbitrary laser-field configurations,
were found with the use of operator techniques in Ref.~\cite{BaMiSt1975}.
Independently, by means of another method, 
the vacuum-polarization tensor was obtained in Ref.~\cite{BeMi1975}.
Both results are in agreement with each other. 
Of course, the scattering, which is beyond Maxwell's classical
electrodynamics, appears due to the polarization of the electron-positron 
vacuum in the presence of a laser field.

\begin{figure}[htb]
\vspace{40pt} \centering
\includegraphics[width=1.0\linewidth]{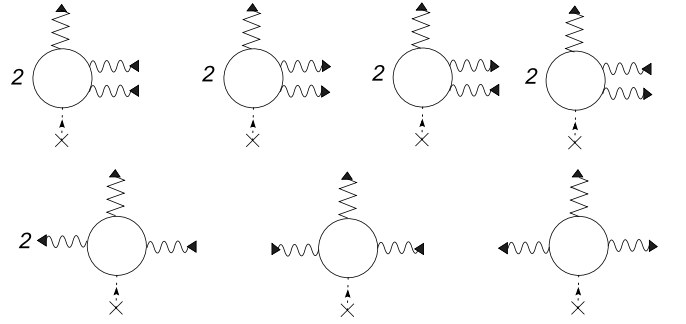}
\begin{picture}(0,0)(0,0)
 \end{picture} 
\caption{\label{diagrams} Diagrams corresponding to the 
amplitude (\ref{T00}), including their multiplicities. 
The solid lines are the electron propagators, the 
dashed line with the cross denotes the Coulomb field, the wavy lines 
correspond to the laser photons, and the 
zigzag line is the induced field.}
\end{figure}

In this article, we use rationalized Gaussian units
with $\hbar=c=1$, and $\epsilon_0 = (4 \pi)^{-1}$,
which implies $e^2 = \alpha$ where $e$ is the electron charge and
$\alpha$ is the fine-structure constant. 
We represent the vector potential of the laser field in the form
\begin{equation}
A_\mu(\phi)=e_{1\mu}\psi_1(\phi)+e_{2\mu}\psi_2(\phi)  \,. 
\end{equation}
Here, $\phi=\varkappa x=\varkappa^0t-{\vec \varkappa}{\vec r}$
is the phase of the laser field at the space-time point 
$(\vec r, t)$, and $\omega = \varkappa^0 = |\vec \varkappa|$ is the 
laser frequency. We use the spacetime metric in 
the Bjorken--Drell convention $g_{\mu\nu} =
{\rm diag}(1,-1,-1,-1)$. The vector $\vec \varkappa$ points into the 
laser propagation direction, and $\psi_{1,2}(\phi)$ are functions which 
characterize the shape of the laser pulse.
We denote the (orthogonal) unit vectors
giving the polarization directions by ${\vec e}_1$ and 
${\vec e}_2$, respectively (${\vec e}^2_1 = {\vec e}^2_1 = 1)$. 
The corresponding four-vectors $e_1 = (0, \vec{e}_1)$ 
and $e_2 = (0, \vec{e}_2)$ satisfy
\begin{equation}
\label{conv1}
\varkappa^2 = 
\varkappa \cdot e_{1} = 
\varkappa \cdot e_{2} =
e_1 \cdot e_2=0\, ,\quad 
e_1^2=e_2^2=-1\,.
\end{equation}
All vector scalar products (involving three or four-vectors)
are denoted by $\cdot$ in this article.
We also use the conventions
\begin{equation}
\label{conv2}
{\vec \nu}=\frac{\vec \varkappa}{\omega}, \qquad
{\vec n}=\frac{{\vec r}}{r}
\end{equation}
for the unit vector ${\vec \nu} = {\vec e}_1 \times {\vec e}_2$
pointing into the propagation direction, and the unit vector pointing 
to a coordinate ${\vec r}$, respectively.

\begin{figure}[htb] 
\includegraphics[width=0.8\linewidth,keepaspectratio=true]{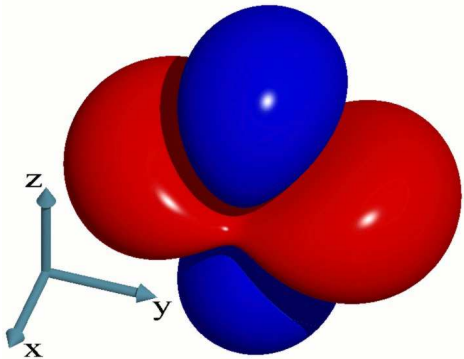}
\caption{\label{isosurf} 
(Color online.)
The familiar rotationally symmetric vacuum-polarization charge
density around an atomic nucleus is severely distorted
in a strong laser field. Here, we plot the additional 
laser-dressed, time-independent contribution to the
renormalized charge density $\delta \rho^R_0(\vec{r})$, 
as induced by the laser field. 
The quantity $\delta \rho^R_0(\vec{r})$ has to be
added to the well-known expressions for the Uehling and the
Wichmann--Kroll parts (for a review see e.g.~\cite{MoPlSo1998}).
The result for $\delta \rho^R_0(\vec{r})$ is obtained after 
Fourier transformation of the 
momentum-space result given in Eq.~(\ref{subtractrho}) below,
into coordinate space. 
Along the vertical axis, we have the propagation
direction of the laser, which we assume to be 
linearly polarized ($\xi_1^2 \neq 0$, $\xi_2^2 = 0$). 
The polarization direction (laser electric
field) is along the $x$-axis.
We plot a characteristic surface given by a constant induced charge density,
namely the surface defined by the relation
$\delta \rho_0(\vec{r}) = \frac{2\alpha}{3\pi}\,\frac{Ze\,m^3}{\pi}\,
\left({\cal E}/{\cal E}_0\right)^2 \, K$, where $K$ is 
a numerical constant given by $K = \pm 1.204$ for the 
plot shown. Two areas with a positive induced charge density
(red, aligned mainly along the $y$ axis, $K = +1.204$) and a negative one
are distinguished (the latter is in blue color and 
located along the propagation axis, $K = -1.204$).
As emphasized in the text, the total induced charge is zero.}
\end{figure}

\begin{figure}
\begin{center}
\includegraphics[width=0.6\linewidth,keepaspectratio=true]{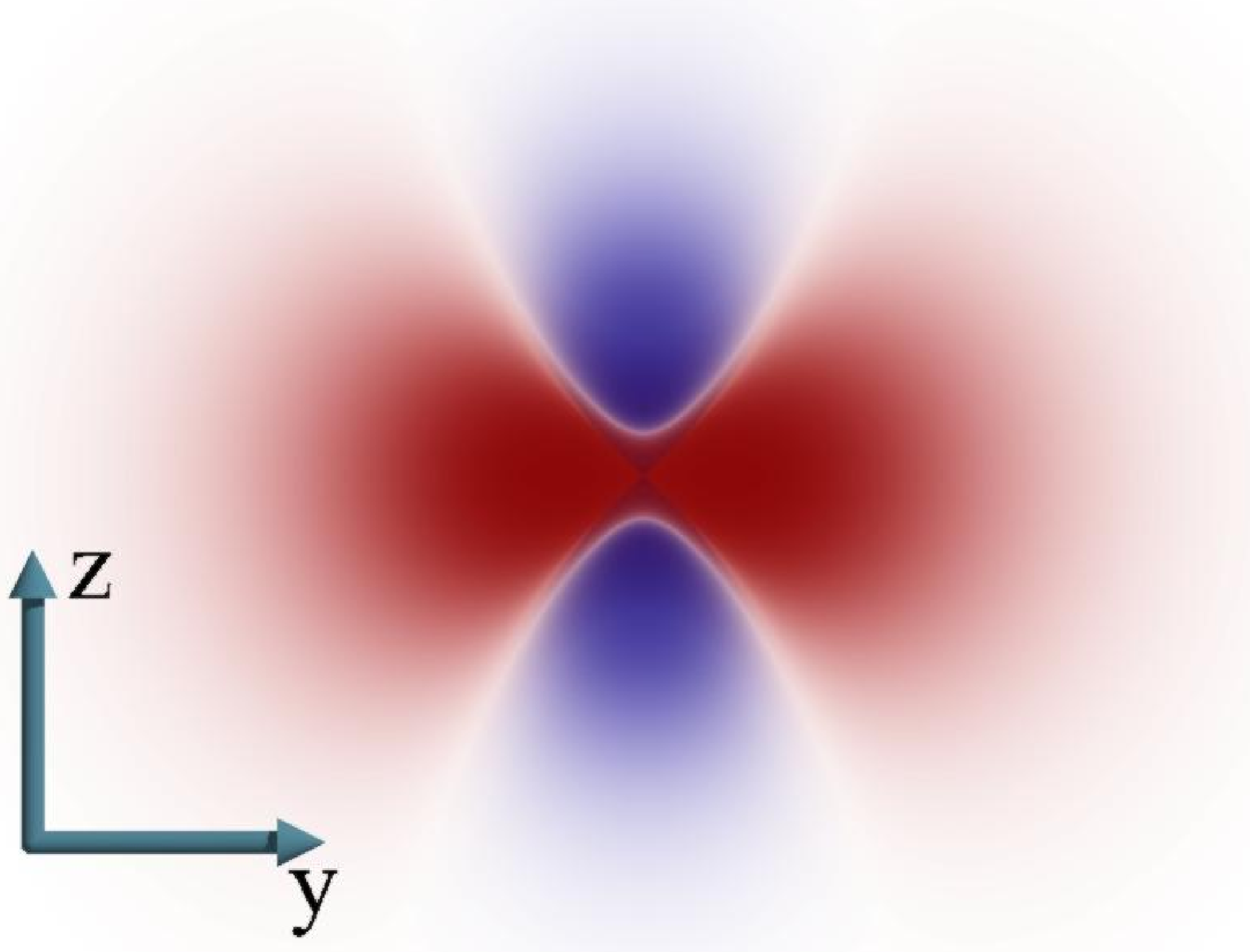}
\centerline{(a)}
\includegraphics[width=0.6\linewidth,keepaspectratio=true]{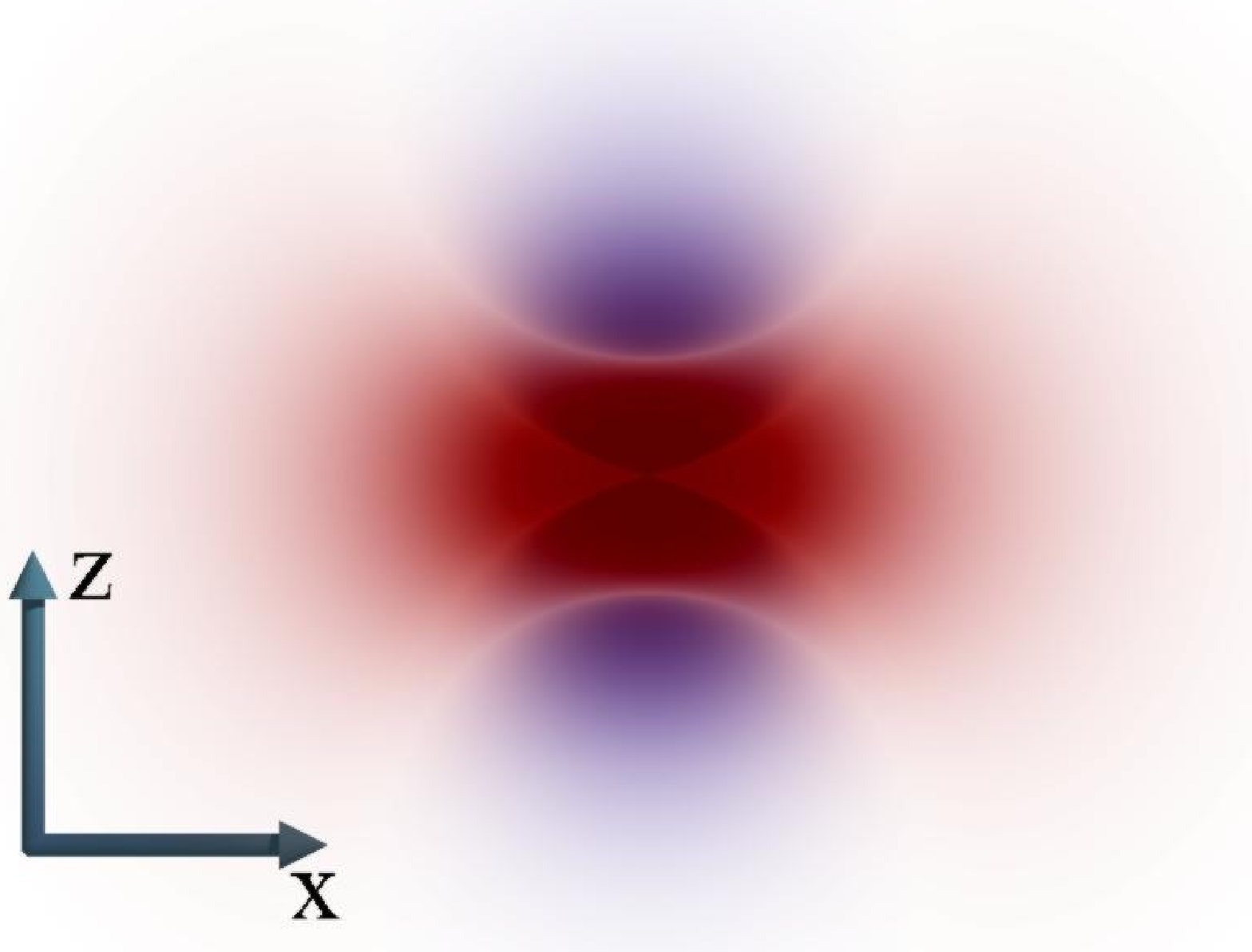}
\centerline{(b)}
\includegraphics[width=0.6\linewidth,keepaspectratio=true]{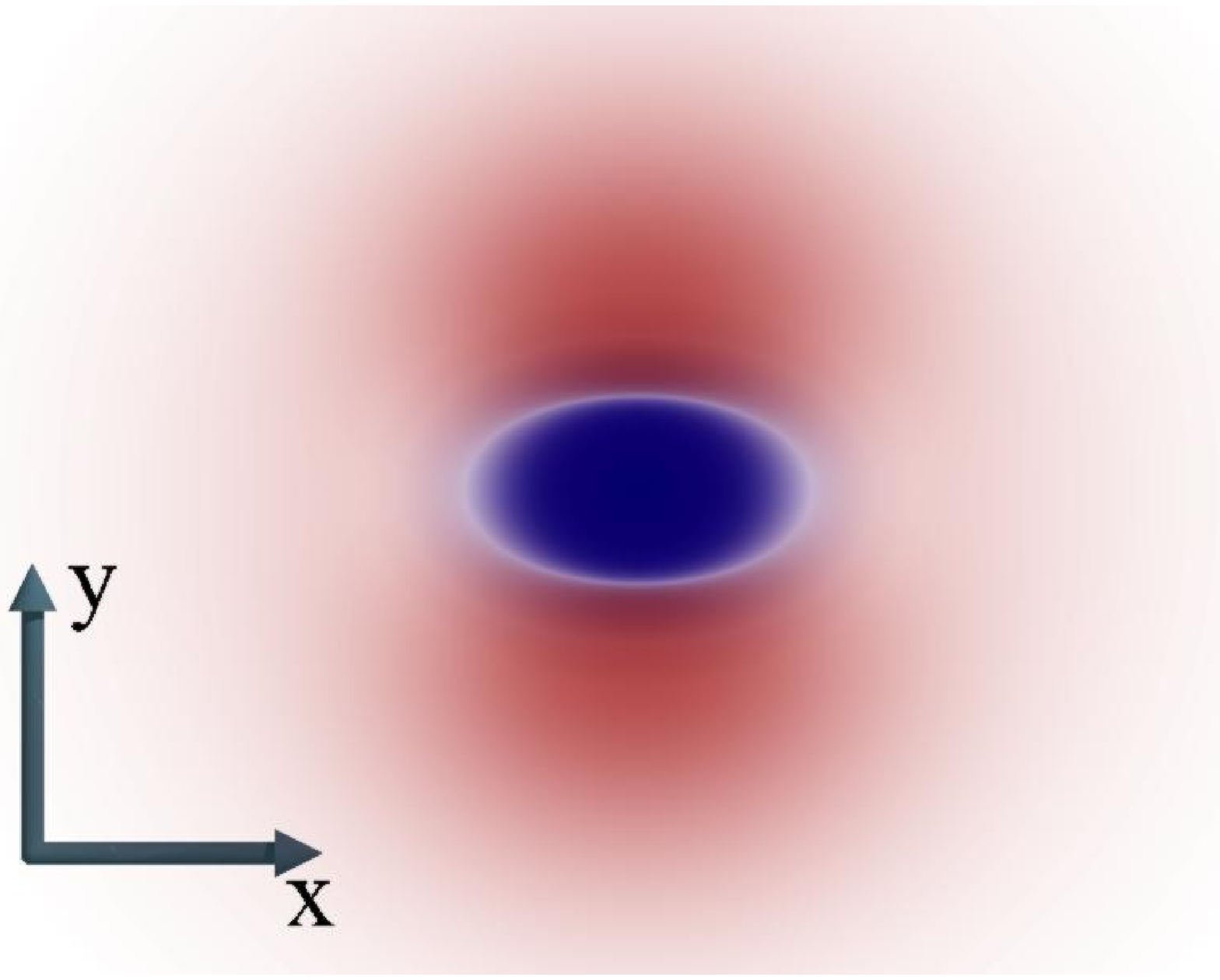}
\centerline{(c)}
\end{center}
\caption{\label{projections} 
(Color online.)
Projections of the laser-induced part of the vacuum-polarization
charge density  looked upon from the $x$, $y$ and $z$ axis,
in Figs.~(a), (b) and (c), respectively.
The conventions are the same as in Fig.~\ref{isosurf}: 
the $z$-axis is the laser propagation direction, and the $x$-axis 
is the polarization axis (laser electric field). 
The picture provides a ray-traced impression
of the charge density, represented as an absorptive 
medium. Red color is used for positive induced charge (areas located 
along the $x$ and $y$ axes), blue color is used for negative
charge (areas located along the $z$ axis).
The plots are as seen by an observer looking at the atomic nucleus from the 
specified directions. This is in contrast to Fig.~\ref{isosurf},
where the isosurface corresponding to a given, constant charge 
density is plotted. In the above graphical representations, 
the fall-off of the induced charge density for large radial 
arguments can be distinguished clearly.}
\end{figure}

In Refs.~\cite{BaMiSt1975,BeMi1975}, the photon scattering amplitude was 
found for arbitrary functions $\psi_{1,2}$. In the current paper, we 
restrict ourselves to the case of monochromatic plane wave with the frequency 
$\omega$, although with allowance for arbitrary (linear, elliptic)
polarizations. In this case, the functions $\psi_1(\phi)$ and 
$\psi_2(\phi)$ take the form
\begin{equation}
\psi_1(\phi) =a_1\cos\phi\, ,\quad  
\psi_2(\phi) =a_2\sin\phi\, ,\quad \varkappa^0=\omega\,. 
\end{equation} 
The scattering amplitude depends on the parameters 
\begin{equation}
\label{parameters}
\xi^2_{1,2} =  \frac{e^2a^2_{1,2}}{m^2}\, , \quad 
\lambda = \frac{\varkappa \cdot k_1}{2m^2} =
\frac{\varkappa \cdot k_2}{2m^2}\, ,
\end{equation} 
where  $m$ is the electron mass.  
The mean square laser field strength ${\cal E}^2$,
averaged over a laser period,
is expressed via the parameters $\xi^2_{1,2}$ 
by the relation
\begin{equation}
\left( \frac{e}{m \, \omega} \right)^2\,{\cal E}^2 =
\frac{1}{2}(\xi_1^2+\xi_2^2)\,,\quad \, .
\end{equation}
 Thus, we may say that 
the parameters $\xi_{1,2}$ 
measure the laser field strength. The parameter $\lambda$, 
by contrast, measures the momentum scales relevant to the 
vacuum-polarization loop (incoming momenta and the four-momentum of the 
laser wave) against the electron mass scale.

The following identity is sometimes useful:
\begin{equation}
\label{identity}
\frac12\, \frac{\omega^2}{m^2} (\xi^2_1 + \xi^2_1) = 
\left(\frac{\cal E}{{\cal E}_0}\right)^2\,.
\end{equation}
Here, ${\cal E}_0$ is the Schwinger critical electric field strength
${\cal E}_0={m^2}/{e}=4.4\times 10^{15}\,\mbox{V\,cm}^{-1}$.

As is well known, the presence of an external field leads
to the existence of an induced current $J^\mu$ due to the vacuum polarization. 
If the external field is a pure Coulomb field, 
then for the induced current we have ${\vec J}=0$, 
and $\rho=J^0$ is well known in this case (see, e.g., \cite{BeLiPi1982}).  
Here, we consider the situation where the external field
is a Coulomb field, but there is, in addition, the laser field, 
which dresses the vacuum-polarization loop.

Our incoming photon is the Coulomb quantum with $k_1=q = (0, {\vec q})$ 
and $\varepsilon_1^\mu=\delta^\mu_0$. Then, $J^\mu=T^{\mu\, 0}(k_2,k_1)$. 
We calculate $J^\mu$ in the leading approximation
in the parameter $Z\alpha$, where $Z$ is the nuclear charge number.
Due to momentum and $C$--parity conservation (Furry theorem), 
we have $k_2=k_1+2l\varkappa$, where $l=0,\pm1,\pm2...$, and 
this fact affords an explanation for the two alternative 
definitions of $\lambda$ in Eq.~(\ref{parameters}).  A noticeable 
contribution to the induced charge density arises at distances $r\simeq 
\lambdabar_{\rm e}=1/m$ , which reciprocally 
corresponds to the momentum of a Coulomb quantum 
$|{\vec q}|\simeq m$.

For all current high-intensity lasers,
the laser photon energy
is small as compared to the electron rest mass energy
($\omega\ll m$), and therefore 
\begin{equation}
\label{assump1}
|\lambda| \sim \frac{|{\vec q}|\,\omega}{m^2} \leq 
\frac{\omega}{m} \ll 1\,.
\end{equation}
Our calculations as reported in the current article
depend on the approximation $\lambda\ll 1$
[see also Eqs.~(2.5), (2.31) and (2.32) of Ref.~\cite{BaMiSt1975}].
All results reported below are given on the first nonvanishing 
order in the expansion in powers of $\omega/m$.
A further condition imposed on our calculations concerns 
the laser intensity parameters $\xi_{1,2}$,
\begin{subequations}
\label{assump2}
\begin{equation}
0 < \xi^2_{1,2} \ll 1/\lambda^2\,,
\end{equation}
which according to Eq.~(\ref{identity}) means that
\begin{equation}
\frac{{\cal E}}{{\cal E}_0} \ll 1\,.
\end{equation}
\end{subequations}
The asymptotics of 
$T^{0\, 0}(k_2,k_1) = T^{0\, 0}(k_2,q)$ under this approximation
are found to 
consist of terms with absorption and emission of one laser photon
each, and of terms with absorption
of two and emission of two laser photons 
(see Fig.~\ref{diagrams}),
\begin{align}
\label{T00}
& T^{0\, 0}(k_2,q)= (2\pi)^4 \,
\left[\delta(q-k_2) \, B^{(0)}({\vec q}) \right.
\nonumber\\
& \left. + \delta(q-k_2-2\varkappa) \, B^{(-)}({\vec q})
+ \delta(q-k_2+2\varkappa) \, B^{(+)}({\vec q})\right]\, .
\end{align}
in the matrix element. Here, the functions $B^{(0)}({\vec q})$
and $B^{(\pm)}({\vec q})$ are given by
\begin{widetext}
\begin{subequations}
\label{BB}
\begin{eqnarray}
B^{(0)}({\vec q})&=&
\frac{\alpha\,\omega^2}{24\pi\,m^2} \, \int\limits_0^1 dv\, (1-v^2)^2\,D^2
\biggl\{-\frac{3}{4}(\xi_1^2-\xi_2^2)(q_1^2-q_2^2)\,
\nonumber\\[1ex]
&& +\frac{(\xi_1^2+\xi_2^2)}{(1-v^2)} \, \left[(q_1^2+q_2^2)(1+v^2+D)+
\frac{({\vec q}^{\,2})^2}{4m^2}(1-v^2)^2D\right]\biggr\}\,,
\\
B^{(\pm)}({\vec q})&=& \frac{\alpha \, \omega^2}{48\pi\,m^2}\,
\int\limits_0^1 dv\, (1-v^2)^2 \, D^2\,
\biggl\{\pm 3\,i\,\xi_1\xi_2\,q_1\,q_2
+\frac{3}{4}(\xi_1^2+\xi_2^2)(q_1^2-q_2^2)
\nonumber\\[1ex]
&& -\frac{(\xi_1^2-\xi_2^2)}{(1-v^2)}\left[(q_1^2+q_2^2)(1+v^2+D)+
\frac{({\vec q}^{\,2})^2}{4m^2}(1-v^2)^2 \, D\right]
\biggr\}\, ,
\end{eqnarray}
\end{subequations}
\end{widetext}
where
\begin{align}
\label{D}
& q_1= {\vec q}\cdot{\vec e}_1,\quad  
q_2={\vec q} \cdot {\vec e}_2,\nonumber\\
& D \equiv D(\vec{q}^{\,2}) = 
\left[1+\frac{{\vec q}^{\,2}(1-v^2)}{4m^2}\right]^{-1}\, .
\end{align}
In the derivation of Eq.~(\ref{BB}), we have neglected higher-order
terms in the expansion in powers of $\omega/m$.
The Fourier transform of the correction to the induced 
charge density due to the laser field, $\delta\rho(\vec q, t)$, is
essentially given by the Fourier transform of $T^{0\, 0}(k_2, q)$. 
It is a sum of time-dependent and time-independent parts: 
\begin{eqnarray}\label{rho}
\delta\rho({\vec q},t)&=&\delta\rho_0({\vec q})+\delta\rho_1({\vec q},t)\, , 
\end{eqnarray}
where
\begin{align}
\label{rho01}
\frac{\delta\rho_0({\vec q})}{Ze} =&
\frac{1}{{\vec q}^{\,2}}B^{(0)}({\vec q})\,,\\
\frac{\delta\rho_1({\vec q},t)}{Ze} =&
\frac{e^{2i\omega t}}{({\vec q}+2{\vec\varkappa})^2}B^{(-)}({\vec q})+
\frac{e^{-2i\omega t}}{({\vec q}-2{\vec\varkappa})^2}\,
B^{(+)}({\vec q}).\nonumber
\end{align}
We now turn to a more specific evaluation of the 
charge density, in various regions of practical interest. 
First, we investigate (in coordinate space) 
the distance $r \sim a_0 = 1/(Z\alpha m)$, which
typically gives rise to the main corrections to the energy
in bound systems ($a_0$ is the Bohr radius). 
For a laser whose frequency is of the 
order of atomic transitions, we have $\omega \sim (Z\alpha)^2 m \ll
1/a_0$, so that the ``interesting'' distance
fulfill $r\ll 1/\omega$, and the corresponding momenta 
are in the range $q\sim 1/r\gg\omega = |\vec\varkappa|$.
We can therefore replace 
\begin{equation}
\label{approx}
({\vec q}\pm2{\vec\varkappa})^2 \to {\vec q}^{\,2}
\end{equation}
in Eq.~(\ref{rho01}). Note that this replacement, in particular,
does not affect the validity of the expressions 
for radial arguments of the order of the Compton wavelength.
Employing the approximation (\ref{approx}),
we obtain from Eqs.~(\ref{BB}), (\ref{rho}) and (\ref{rho01}) 
\begin{subequations}
\begin{align}
\label{rhoas0}
& \delta\rho_0({\vec q}) =\frac{\alpha(Ze)\omega^2}{24\pi\,m^2{\vec q}^{\,2}}
\int\limits_0^1 dv\, (1-v^2)^2 \, D^2
\nonumber\\
& \times \biggl\{-\frac{3}{4}(q_1^2-q_2^2)
(\xi_1^2-\xi_2^2)\,
+\left[\frac{(q_1^2+q_2^2)}{(1-v^2)}(1+v^2+D)\right.
\nonumber\\
& \quad \left. 
+ \frac{({\vec q}^{\,2})^2}{4m^2}(1-v^2)\,D\right](\xi_1^2+ \xi_2^2)\biggr\}\,,
\end{align}
and
\begin{align}
\label{rhoas1}
& \delta\rho_1({\vec q},t) =
\frac{\alpha(Ze)\omega^2}{24\pi\,m^2{\vec q}^{\,2}}
\int\limits_0^1 dv\, (1-v^2)^2 \, D^2
\\
& \times \biggl\{\frac{3}{4}(q_1^2-q_2^2)
(\xi_1^2\, +\xi_2^2)\cos(2\omega t)
\nonumber\\
& \quad -\left[\frac{(q_1^2+q_2^2)}{(1-v^2)}(1+v^2+D)+
\frac{({\vec q}^{\,2})^2}{4m^2}(1-v^2)\,D\right]
\nonumber\\
& \qquad \times (\xi_1^2- \xi_2^2)\cos(2\omega t)
+3\xi_1\,\xi_2\,q_1\,q_2\,\sin(2\omega t) \biggr\}\,.\nonumber
\end{align}
\end{subequations}
Although the expression for the vacuum polarization tensor
in Eqs.~(2.5), (2.31) and (2.32) of Ref.~\cite{BaMiSt1975}
is in principle already renormalized, it is still
necessary to carry out an additional finite subtraction,
namely, to subtract the asymptotics of the induced charge
at $|{\vec q}|\rightarrow 0$. 
This is akin to the additional subtractions that have 
to be carried out in the calculation of the Wichmann--Kroll
correction. The renormalization condition in the on-mass-shell scheme,
for the Coulomb field is that two distant observers should 
observe exactly the Thomson scattering cross section, which 
implies that the total induced charge has to be equal to zero
(an eludicating discussion surrounds Eq.~(7.18) of Ref.~\cite{ItZu1980}). 
In addition, our subtraction provides a finite quadrupole moment, 
corresponding to the induced charge distribution. Because
$\delta\rho_1({\vec q},t)\rightarrow 0$ at $|{\vec q}|\rightarrow 0$ 
[see Eq.(\ref{rho01})], it is only the time-independent part 
$\delta\rho_0({\vec q})$ which receives an additional,
finite renormalization. 

The subtraction leads to the renormalized time-independent 
component of the charge density $\delta\rho_0^R({\vec q})$,
\begin{align}
\label{subtractrho}
& \delta\rho_0^R({\vec q})=
\delta\rho_0({\vec q})-\frac{\alpha(Ze)\omega^2}{60\pi\,m^2{\vec q}^{\,2}}
\nonumber\\
& \times \left[\frac{11}{3}\,
(\xi_1^2+ \xi_2^2)(q_1^2+q_2^2)-(\xi_1^2- \xi_2^2)(q_1^2-q_2^2)\right]\,.
\end{align}
The renormalized charge density $\delta\rho_0^R({\vec r})$,
excluding the Dirac delta contribution, is represented 
graphically in Figs.~\ref{isosurf} and~\ref{projections}.
The full renormalized $\delta\rho^R({\vec q},t)$ 
(including the time-dependent term) is
\begin{equation}
\delta\rho^R({\vec q},t)=\delta\rho_0^R({\vec q})+\delta\rho_1({\vec q},t)\, .
\end{equation}
The potential $\Phi({\vec r},t)$ in the Lorentz gauge satisfies the equation
\begin{equation}
\label{lorj0}
\frac{\partial^2}{\partial t^2} \, \Phi({\vec r},t) -
\vec\nabla^2 \Phi({\vec r},t) = 
4 \pi \, \delta\rho^R({\vec r},t)\, .
\end{equation}
\begin{widetext}
Therefore, the Fourier transform of the potential, 
which corresponds to the density $\delta\rho^R({\vec q},t)$, is
\begin{eqnarray}
\label{Phiqt}
\Phi({\vec q},t)&=&\frac{4\pi}{{\vec q^2}}\,
\delta\rho_0^R({\vec q})+\frac{4\pi}{{\vec q}^{\,2}-4 \omega^2}\,
\delta\rho_1({\vec q},t)\nonumber\\
&=&\frac{4\pi}{{\vec q^2}}\,\delta\rho_0^R({\vec q})+
4\pi(Ze)\left[ \frac{e^{2i\omega t}B^{(-)}({\vec q})}{
({\vec q}^{\,2}-4\omega^2+i0)({\vec q}+2{\vec\varkappa})^2}+
\frac{e^{-2i\omega t}B^{(+)}({\vec q})}{({\vec q}^{\,2}-4\omega^2-i0)
({\vec q}-2{\vec\varkappa})^2}\right]  \, .
\end{eqnarray}
Here the regularization of the denominators, ($\pm i0$), 
corresponds to the boundary condition of radiation (see below).
It is convenient to present the potential in coordinate 
space as a sum of time-dependent and
time-independent parts:
\begin{equation}\label{phi01}
\Phi({\vec r},t)=\Phi_0({\vec r})+\Phi_1({\vec r},t)\, .
\end{equation}
By Fourier transform of Eq.~(\ref{Phiqt}), we thus obtain at 
a distance $r\ll 1/\omega$ from the nucleus,
\begin{subequations}
\label{Phi}
\begin{eqnarray}
\label{Phi0}
\Phi_0({\vec r})&=&\frac{\alpha(Ze)\omega^2}{48\pi\,m}\biggl\{
\left[F_1(mr)+F_2(mr)\left(1-3({\vec \nu} \cdot {\vec n})^2\right)\right]
(\xi_1^2+ \xi_2^2)
\nonumber\\
&&-F_3(mr) 
\left[ ({\vec n} \cdot {\vec e}_1)^2
-({\vec n} \cdot {\vec e}_2)^2\right](\xi_1^2- \xi_2^2)\biggr\}\, ,\\
\label{Phi1}
\Phi_1({\vec r},t)&=&-\frac{\alpha(Ze)\omega^2}{48\pi\,m}\biggl\{
\left[\left(F_1(mr)+\frac{88}{45mr}\right)+\left(F_2(mr)-
\frac{22}{45mr}\right)\left(1-3({\vec \nu}{\vec n})^2\right)\right]
(\xi_1^2- \xi_2^2)\cos(2\omega t) 
\nonumber\\
&&-\left(F_3(mr)-\frac{2}{5mr}\right)\left[ (\xi_1^2
+\xi_2^2)\cos(2\omega t)
\left[ ({\vec n} \cdot {\vec e}_1)^2
-({\vec n} \cdot {\vec e}_2)^2\right]
+ 4\xi_1\xi_2\sin(2\omega t)
({\vec n} \cdot {\vec e}_1) ({\vec n} \cdot {\vec e}_2)\right]\biggr\},
\end{eqnarray}
\end{subequations}
where we note the conventions in Eqs.~(\ref{conv1}) and~(\ref{conv2}).
In this formula, 
\begin{subequations}
\label{F123}
\begin{eqnarray}
F_1(mr)&=&\int\limits_0^1 dv\, 
(1-v^2)^{1/2}\left\{\frac{1}{2}(1-v^2)(1+2y)-
\frac{4}{3y} \, \left[(1+v^2)(1+y)+1
+\frac{5}{4}y+\frac{1}{2}y^2\right]\right\}e^{-2y}\,,
\\
F_2(mr)&=&\int\limits_0^1 dv\, (1-v^2)^{1/2} \,
\biggl\{ \frac{1}{3y}\left[(1+v^2)(1+y)+1
+\frac{5}{4}y+\frac{1}{2}y^2\right]e^{-2y}
\nonumber\\
&& + \frac{1+v^2}{y^3}\left[1-(1+y)(1+y+y^2)e^{-2y}\right]
+ \frac{3}{2y^3}\left[1-(1+2y+2y^2+\frac{7}{6}y^3
+ \frac{1}{3}y^4)e^{-2y}\right]\biggr\}\, ,
\\
F_3(mr)&=&\frac{9}{4}\,
\int\limits_0^1 dv\,\frac{(1-v^2)^{3/2}}{y^3}
\left[1-(1+y)(1+y+\frac{2}{3}y^2) \, e^{-2y}\right]\,,\nonumber\\
y& \equiv& y(r,v) = \frac{mr}{\sqrt{1-v^2}}\,.
\end{eqnarray}
\end{subequations}
\end{widetext}
In principle, the expressions (\ref{Phi}) gives the 
complete answer for the induced potential, which is necessary for the 
calculation of the energy shift.
We reemphasize that the only additional approximation employed was 
$\lambda \ll 1$, which is 
easily fulfilled for current and planned high-intensity 
laser facilities.
Note that in the particular case of the circular polarized laser field 
($\xi_1=\pm\xi_2=\xi$), the expression for the induced potential becomes 
essentially simpler.

%
%
\subsection{Asymptotics of the induced potential}
\label{asymppotential}

Let us consider the behaviour  of  the potential (\ref{Phi}) at 
large and small distances. The functions $F_i(\rho)$ have the 
following asymptotics at $\rho=mr\gg 1$:
\begin{eqnarray}
\label{asymp1}
F_1(\rho)\simeq\frac{\sqrt{\pi\rho}}{6}e^{-2\rho}\,,\;
F_2(\rho)\simeq\frac{148}{105\rho^3}\, ,\;
F_3(\rho)\simeq\frac{36}{35\rho^3}.
\end{eqnarray}
We can thus safely neglect the exponentially decaying $F_1(\rho)$-term against 
the polynomial behaviour of $F_2$ and $F_3$ for $r \gg 1/m$. 
We recall that the representation in Eqs.~(\ref{Phi}) and~(\ref{F123}) 
is valid under the (additional) condition $r \ll 1/\omega$.
Thus, we write the asymptotics of the potential at $1/\omega \gg r\gg 1/m$ 
in the form
\begin{eqnarray}
\label{phir}
\Phi({\vec r},t)&\simeq & 
\frac{n^i\,n^jQ^{ij}}{2r^3}+\frac{1}{r}f({\vec n},t)\,,\\
f({\vec n},t)&=& -\frac{\alpha(Ze)\omega^2}{360\pi\,m^2}\biggl\{
11\left[1+({\vec \nu} \cdot {\vec n})^2\right]
(\xi_1^2- \xi_2^2)\cos(2\omega t) 
\nonumber\\
&&+3(\xi_1^2
+\xi_2^2)\cos(2\omega t)
\left[ ({\vec n} \cdot {\vec e}_1)^2
-({\vec n} \cdot {\vec e}_2)^2\right]
\nonumber\\
&& + 12\xi_1\xi_2\sin(2\omega t)\,
({\vec n} \cdot {\vec e}_1) \, ({\vec n} \cdot {\vec e}_2)\biggr\}\, .
\end{eqnarray}
The first term on the right-hand side of Eq.~(\ref{phir}), 
proportional to $1/r^3$, can be associated in a natural
way with a charge distribution that originates from an
oscillatory quadrupole moment,
\begin{widetext}
\begin{eqnarray}\label{Q}
Q^{ij}&=&\frac{3\alpha(Ze)\omega^2}{35\pi\,m^4}\biggl\{
\frac{37}{27}\left(\delta^{ij}-3\nu^i\nu^j\right)
\left[\xi_1^2\sin^2(\omega t)+ \xi_2^2\cos^2(\omega t)\right] 
\nonumber\\
&&-(e_1^i e_1^j-e_2^i e_2^j)
\left[\xi_1^2\sin^2(\omega t)- \xi_2^2\cos^2(\omega t)\right]
-(e_1^i e_2^j+e_2^i e_1^j)
\xi_1\xi_2\sin(2\omega t)\biggr\}\, .
\end{eqnarray}
\end{widetext}
The second term on the right-hand side of (\ref{phir}), 
proportional to $1/r$, reflects the 
existence of radiation in the system,
but cannot be used for a valid description of the 
radiation.
The reason is that the derivation of the asymptotics (\ref{phir}) 
was carried out at distances $1/m\ll r\ll 1/\omega$. This 
is insufficient for the consideration of radiation, 
which by definition is only a well-defined process 
at distances much larger than the wavelength of the emitted light.
For the calculation of the radiation, it is necessary to consider  
distances $r\gg 1/\omega$ (see Sec.~\ref{photoncurrent}) below.

As was already pointed out, for the purpose of calculating  
the energy shift, it is necessary to know the time-independent part 
of the potential (for the oscillatory terms,
the time-averaging is applied over a full period of the driving laser).
The time-averaged part of the quadrupole moment has the form:
\begin{align}\label{Q0}
Q_0^{ij} =& \frac{3\alpha(Ze)\omega^2}{70\pi\,m^4}\biggl\{
\frac{37}{27}\left(\delta^{ij}-3\nu^i\nu^j\right)
(\xi_1^2+ \xi_2^2) 
\nonumber\\
& -(e_1^i e_1^j-e_2^i e_2^j)(\xi_1^2- \xi_2^2)\biggr\}\, .
\end{align}

At small distances from the nucleus,
$r\ll 1/m$, we get a complete different asymptotic behaviour for the 
functions $F_i(\rho)$,
\begin{eqnarray}
\label{asymp2}
F_1(\rho)\simeq-\frac{88}{45\rho}\,,\quad
F_2(\rho)\simeq\frac{22}{45\rho}\, ,\quad 
F_3(\rho)\simeq\frac{2}{5\rho}\,.
\end{eqnarray} 
For the potential, this implies
\begin{eqnarray}
\Phi({\vec r},t)&\simeq&-\frac{\alpha(Ze)\omega^2}{360\pi\,m^2\, r }
\biggl\{ 11\left[1+({\vec \nu} \cdot {\vec n})^2\right](\xi_1^2+ \xi_2^2) 
\nonumber\\
&&+3 \left[({\vec n} \cdot {\vec e}_1)^2-
({\vec n} \cdot {\vec e}_2)^2\right](\xi_1^2- \xi_2^2)\biggr\}\, .
\end{eqnarray}
We see that the leading small-distance asymptotics is independent of time. 

%
%
\subsection{Application: energy shift of hydrogenic levels}
\label{energypotential}

Let us consider  the correction 
$\delta E=\left<\,e\Phi_0({\vec r})\,\right>$ to hydrogenic energy levels, 
corresponding to the potential $\Phi_0({\vec r})$, Eq.(\ref{Phi}).
For $n\,s_{1/2}$-states, 
only the spherically symmetric part of this potential
gives a nonzero contribution.
It is proportional to the function $F_1(mr)$.
Because the function $F_1(mr)$ decreases exponentially at $mr\gg 1$, we 
can replace in the matrix element the wave function $\psi(r)$ by $\psi(0)$, 
because it has the electron Compton wavelength is much 
smaller than the characteristic scale $a_0=1/(Z\alpha m)\gg 1/m$.   

With the considerations, the energy shift for $S$
states can be easily calculated, and we obtain 
[see Eqs.~(\ref{identity}),~(\ref{Phi})~and~(\ref{asymp1})]
\begin{eqnarray}
\label{es}
\delta E(nS)&=&
-\frac{8\alpha(Z\alpha)^4\omega^2}{135\pi\,n^3m}(\xi_1^2+\xi_2^2)
\nonumber\\
&=&
-\frac{16\alpha(Z\alpha)^4\,m}{135\pi\,n^3}
\left(\frac{{\cal E}}{{\cal E}_0}\right)^2\,  .
\end{eqnarray}
Therefore, due to the factor  ${\cal E}/{\cal E}_0$, 
in all realistic cases the correction obtained is small.

Let us consider the corrections to the energy level with the orbital 
angular momentum $l >0$. In the general case of an elliptically polarized 
wave ($\xi_1\neq\xi_2$) , the projection $j_z$ of the total angular 
momentum on the propagation direction of the laser
$\vec\nu$ is not conserved. 
This implies a nonvanishing nondiagonal matrix element of the 
potential $\Phi_0({\vec r})$ for $\Delta j_z=\pm 2$. 
For $j\ge 3/2$ (where $j$ is the 
total electron angular momentum),
it is necessary to solve a more complicated secular equation
in order to find the eigenvalues in the presence of the 
laser-dressed vacuum polarization. For 
simplicity, we restrict ourselves to the case of a circular polarized 
laser field, where the mentioned problem does not appear,
and all matrix elements remain diagonal. 

For $l\ne 0$, the main 
contribution to the integral for the matrix element $\Phi_0({\vec r})$
comes from distances of the order of the 
Bohr radius, $r\sim a_0 \gg 1/m$, and is generated by the 
quadrupole potential. Therefore, we can substitute to the 
matrix element the asymptotics of the potential 
$\Phi_0({\vec r})\simeq Q^{i\,j}_0n^in^j/(2r^3)$
with the quadrupole moment [see Eqs. (\ref{phir}) and (\ref{Q0})] 
being equal to
\begin{eqnarray}\label{QC}
Q^{ij}_0&=&\frac{37\alpha(Ze)\omega^2\xi^2}{315\pi\,m^4}
\left(\delta^{ij}-3\nu^i\nu^j\right) 
\end{eqnarray}
for the case of circular polarization of the laser field. Using 
the relations~\cite{BeLiPi1982}
\begin{align}
& \left<\frac{1}{r^3}\right> = \frac{(Z\alpha m)^3}{n^3\,l(l+1)(l+1/2)}\,,
\nonumber\\
& \int d\Omega\, |Y_{lm}|^2(1-3\cos^2\theta)=
\frac{3m^2-l(l+1)}{(2l+3)(l-1/2)}\,,
\end{align}
where $l$ is the orbital angular momentum,
we obtain the following shift of energy level with $l\ne 0$ 
\begin{eqnarray}\label{ep}
\delta E_{jl}&=&\frac{37\alpha(Z\alpha)^4m}{315\pi\,n^3}\,
\left(\frac{{\cal E}}{{\cal E}_0}\right)^2\,
\frac{3j_z^2-j(j+1)}{l(l+1)(2l+1)^2}
\nonumber\\
& & \times \left[\frac{\delta_{j-1/2,\,l}}{j+1}+
\frac{\delta_{j+1/2,\,l}}{j}\right]\,.
 \end{eqnarray}
In rewriting the result in terms of the total rather than orbital
angular momentum of the electron, we have used the explicit 
representation of the spinor spherical biharmonics~\cite{BeLiPi1982}. 
It is interesting that the magnitude of the correction for $l\ne 0$ is 
of the same order as that for $l=0$.  
This is in sharp contrast to the ``normal'' vacuum polarization
in a Coulomb field, where the dominant effect is given 
by a potential of the form $-(1/15)\,\alpha\,Z\alpha\,m^2\,\delta(\vec{r})$,
which has a nonvanishing expectation value only for $S$ states.

%
%
\section{INDUCED CURRENT AND VECTOR POTENTIAL}
\label{current}

%
%
\subsection{From the current to the vector potential}
\label{calccurrent}

In a pure Coulomb field acting as an external ``condition'',
or source, it is 
well known that the spatial components of the induced current
vanish completely.
The situation changes drastically for an additional laser field.
Then, the spatial components of the induced current no longer
vanish, and that constitutes an additional effect
supplementing the vacuum-induced laser-dressed
charge density~$\delta \rho$, which exists in both cases.
We discuss the Fourier transform of the induced nonvanishing 
current density ${\vec J}({\vec q},t)$. We employ analogous steps as in the 
derivation of Eq.~(\ref{T00}).

We define the following vector to represent 
the mixed spatial-temporal
components of the vacuum-polarization tensor,
\begin{equation}
T^{i\, 0}(k_2,k_1)\equiv {\vec T}^i(k_2,k_1)\,.
\end{equation}
In analogy to Eq.~(\ref{Phiqt}), we find the following 
structure,
\begin{align}
& {\vec T}(k_2,k_1) = 
(2\pi)^4\bigl[\delta(k_1-k_2){\vec G}^{(0)}({\vec q}) 
\nonumber\\
& +\delta(k_1-k_2-2\varkappa){\vec G}^{(-)}({\vec q})+
\delta(k_1-k_2+2\varkappa){\vec G}^{(+)}({\vec q})\bigr]\,.
\end{align}
The functions ${\vec G}^{(0)}({\vec q})$ and ${\vec G}^{(\pm)}({\vec q})$
are found to read 
\begin{widetext}
\begin{align}
\label{GG}
{\vec G}^{(0)}({\vec q}) =&
\frac{\alpha\,\omega^2}{24\pi\,m^2}\int\limits_0^1 dv\, (1-v^2)^2D^2
\biggl\{\frac{\xi_1^2+\xi_2^2}{1-v^2}(2+v^2D) \,
\left[{\vec q}\times\left[{\vec\nu}\times{\vec q}\right]\right]-
\frac{3}{4}(\xi_1^2-\xi_2^2) \,
\left[{\vec q}\times\left[{\vec\nu}\times({\vec q}_1-
{\vec q}_2)\right]\right]\biggr\}\,,
\nonumber \\
{\vec G}^{(\pm)}({\vec q}) =&
\frac{\alpha\,\omega^2}{48\pi\,m^2}\,
\int\limits_0^1 dv\, (1-v^2)^2D^2
\biggl\{\frac{3}{4}(\xi_1^2+\xi_2^2)\,
\left[\pm 2\omega({\vec q}_1-{\vec q}_2)+
\left[{\vec q}\times\left[{\vec\nu}\times({\vec q}_1-
{\vec q}_2)\right]\right]\right]
\nonumber\\
&-\frac{\xi_1^2-\xi_2^2}{1-v^2}(2+v^2D)\,
\left[\pm 2\omega({\vec q}_1+{\vec q}_2)+
\left[{\vec q}\times\left[{\vec\nu}\times{\vec q}\right]\right]\right]\pm
\frac{3}{2}\, i \, \xi_1\xi_2 \, 
\left[(\pm 2\omega{\vec\nu}-
{\vec q})\times({\vec q}_1-{\vec q}_2) \right] \,
\left[ {\vec \nu} \cdot \left( {\vec e}_1 \times {\vec e}_2 \right) \right] 
\biggr\}\, .
\end{align}
\end{widetext}
Here, $D$ is defined as in Eq.~(\ref{D}), 
and we recall the conditions
(\ref{assump1}) and (\ref{assump2}) which apply 
to all results reported in this paper.

Although ${\vec \nu} = {\vec e}_1 \times {\vec e}_2$ and 
${\vec \nu}^2 = 1$ in a specified 
coordinate system [see Eq.~(\ref{conv2})], 
it is still useful to indicate the factor 
\begin{equation}
{\vec \nu} \cdot 
\left( {\vec e}_1 \times {\vec e}_2 \right).
\end{equation}
The reason is that under a parity transformation,
${\vec G}^{(\pm)}({\vec q})$ should transform like a vector,
and the expression $(\pm 2\omega{\vec\nu}-
{\vec q})\times({\vec q}_1-{\vec q}_2)$ would otherwise transform 
like an axial vector.
In Eq.~(\ref{GG}), $D$ is defined as in Eq.~(\ref{D}), and
${\vec q}_i = {\vec e}_i({\vec e}_i{\vec q})$.
The result is valid under the conditions 
(\ref{assump1}) and (\ref{assump2}).

In analogy to Eq.~(\ref{rho}), 
we represent the induced current as 
a sum of the time-independent and time-dependent parts:
\begin{eqnarray}
{\vec J}({\vec q},t)= {\vec J}_0({\vec q})+
{\vec J}_1({\vec q},t)\, ,
\end{eqnarray} 
where
\begin{align}
\frac{{\vec J}_0({\vec q})}{Ze} =&
\frac{{\vec G}^{(0)}({\vec q})}{{\vec q}^{\,2}} \,, \\
\frac{{\vec J}_1({\vec q},t)}{Ze} =& 
\frac{e^{2i\omega t}}{({\vec q}+2{\vec\varkappa})^2}{\vec G}^{(-)}({\vec q})
+\frac{e^{-2i\omega t}}{({\vec q}-
2{\vec\varkappa})^2}{\vec G}^{(+)}({\vec q})\,.\nonumber
\end{align} 

In analogy to Eq.~(\ref{subtractrho}), 
we now have to subtract from ${\vec J}_0({\vec q})$ its
asymptotics at $|{\vec q}|\rightarrow 0$. A time-independent 
component of ${\vec J}_0({\vec q})$, constant in momentum 
space and therefore local in coordinate space, would 
otherwise lead to a magnetic-monopole field.
In order to preserve the nonexistence of a magnetic monopole,
we introduce a finite subtraction $\tilde{\vec J}({\vec q})$ 
in the time-like component of the current ${\vec J}_0({\vec q})$,
so that the renormalized current ${\vec J}^R({\vec q},t)$ has the form
\begin{equation}
\label{subtractJ}
{\vec J}^R({\vec q},t)={\vec J}_0^R({\vec q})+{\vec J}_1({\vec q},t)=
[{\vec J}_0({\vec q})-\tilde{\vec J}({\vec q})]+ {\vec J}_1({\vec q},t)\,.
\end{equation}
Here, the finite renormalization $\tilde{\vec J}({\vec q})$ has the form
\begin{align}
\tilde{\vec J}({\vec q}) =&
\frac{\alpha(Ze)\omega^2}{180\pi\,m^2{\vec q}^{\,2}}
\biggl\{11(\xi_1^2+\xi_2^2)
\left[{\vec q}\times\left[{\vec\nu}\times{\vec q}\right]\right]
\nonumber\\
& -3(\xi_1^2-\xi_2^2)\left[{\vec q}\times\left[{\vec\nu}\times
({\vec q}_1-{\vec q}_2)\right]\right]\biggr\}\, .
\end{align} 

In order to infer the induced vector potential ${\vec A}({\vec r},t)$,
one now has to solve the equation
\begin{equation}
\label{lorji}
\frac{\partial^2}{\partial t^2}\, {\vec A}({\vec r},t) - 
\vec\nabla^2 {\vec A}({\vec r},t)=4\pi \, {\vec J}^R({\vec r},t).
\end{equation}
In Fourier space, the solution is immediate,
and ${\vec A}({\vec q},t)$ is found to be
\begin{align}
{\vec A}({\vec q},t)=&\frac{4\pi}{{\vec q}^{\,2}}{\vec J}_0^R({\vec q})
\nonumber\\
& +4\pi(Ze)\left[
\frac{e^{2i\omega t}{\vec G}^{(-)}({\vec q})}{({\vec q}^{\,2}-4\omega^2+i0)
 ({\vec q}+2{\vec\varkappa})^2}\right.
\nonumber\\
& \left. +\frac{e^{-2i\omega t}{\vec G}^{(+)}({\vec q})}
  {({\vec q}^{\,2}-4\omega^2-i0)({\vec q}-2{\vec\varkappa})^2}\right]\, .
\end{align} 

%
%
\subsection{Current and vector potential in various asymptotic regions}
\label{asympvectorpot}

We first consider the momentum range $|{\vec q}|\gg \omega$, which is
relevant for atomic physics phenomena. In coordinate space,
this condition corresponds to distances
$ r\ll 1/\omega$. We find 
\begin{widetext}
\begin{eqnarray}
\lefteqn{{\vec J}^R({\vec q},t)=
\frac{\alpha(Ze)\omega^2}{16\pi\,m^2{\vec q}^{\,2}}
\int\limits_0^1 dv\, (1-v^2)^2D^2
\biggl\{\frac{4(2+v^2D)}{3(1-v^2)}
[\xi_1^2\sin^2(\omega t)+
\xi_2^2\cos^2(\omega t)]
\left[{\vec q}\times\left[{\vec\nu}\times{\vec q}\right]\right]}
\nonumber \\
&&-[\xi_1^2\sin^2(\omega t)-\xi_2^2\cos^2(\omega t)]
\left[{\vec q}\times\left[{\vec\nu}\times({\vec q}_1-
{\vec q}_2)\right]\right]
-\xi_1\xi_2\sin(2\omega t)
\left[{\vec q}\times({\vec q}_1-{\vec q}_2)\right]
\left[ {\vec \nu} \cdot \left( {\vec e}_1 \times {\vec e}_2 \right) \right] 
\biggr\}
-\tilde{\vec J}({\vec q})\, ,
\end{eqnarray}
where we recall the definition of $D$ in Eq.~(\ref{D}).
The renormalized current ${\vec J}^R({\vec q},t)$ contains both 
time-independent components as well as time-dependent components
(the latter vanish when averaged over a full laser period).

We now use the additional approximation
$r\gg 1/m$. We find that 
the asymptotics of the induced current 
density in the range $1/\omega\gg r\gg 1/m$ are given by 
the purely time-dependent component
\begin{eqnarray}
\label{asJ}
\lefteqn{ {\vec J}^R({\vec r},t) \simeq {\vec J}_1({\vec r},t)
\simeq\frac{\alpha(Ze)\omega^2}{80\pi^2\,m^2r^3}
\left\{\frac{11}{9}(\xi_1^2-\xi_2^2)
\left[{\vec\nu}-3{\vec n}({\vec\nu} \cdot {\vec n})\right]\cos(2\omega t)
\right.} \nonumber \\
&& \left. -(\xi_1^2+\xi_2^2)
\left[{\vec\nu}\left(n_1^2-n_2^2\right)-
({\vec n}_1-{\vec n}_2) ({\vec\nu} \cdot {\vec n})\right]\cos(2\omega t) 
+2\xi_1\xi_2\sin(2\omega t)
\left[{\vec n}\times({\vec n}_1-{\vec n}_2)\right]
\left[ {\vec \nu} \cdot \left( {\vec e}_1 \times {\vec e}_2 \right) \right] 
\right\}\,.
\end{eqnarray}
%
Here, we have used the conventions
\begin{eqnarray}
n_{i}=({\vec e}_{i} \cdot {\vec n})\quad ,\quad 
{\vec n}_{i}=({\vec e}_{i} \cdot {\vec n}){\vec e}_{i}\, .
\end{eqnarray}
Thus, ${\vec J}_1({\vec r},t)\propto 1/r^3$ in the
range $1/\omega \gg r\gg 1/m$. The 
time-independent current ${\vec J}_0^R({\vec r})$ at distances 
$r\gg 1/m$ diminishes exponentially, 
in a way similar to the time-independent 
induced charge density $\rho_0^R({\vec r})$,
and therefore is omitted in Eq.~(\ref{asJ}).

Again, in the range $1/m\ll r\ll 1/\omega$, we now calculate
the time-independent and time-dependent components of the 
vector potential, ${\vec A}_0({\vec r})$ and  ${\vec A}_1({\vec r},t)$, 
respectively. 
These correspond to the currents ${\vec J}_0^R({\vec r})$ and 
${\vec J}_1({\vec r},t)$. The result reads
\begin{subequations}
\begin{align}
\label{A0}
{\vec A}_0({\vec r}) =
\frac{\alpha(Ze)\omega^2}{140\pi\,m^4r^3}
\biggl\{\frac{31}{9}(\xi_1^2+\xi_2^2)
\left[{\vec\nu}-3{\vec n}({\vec\nu} \cdot {\vec n})\right]
- 3(\xi_1^2-\xi_2^2)\left[{\vec\nu}({n_1^2}-{n_2^2})-
({\vec n}_1-{\vec n}_2)({\vec\nu} \cdot {\vec n})\right]\biggr\}\, ,
\end{align}
and
\begin{align}
\label{A1}
{\vec A}_1({\vec r},t)=&
-\frac{\alpha(Ze)\omega^2}{360\pi \, m^2r}
\biggl\{11(\xi_1^2-\xi_2^2)\left[{\vec\nu}+
{\vec n}({\vec\nu} \cdot {\vec n})\right]\cos(2\omega t)
+ 3(\xi_1^2+\xi_2^2)\left[{\vec\nu}({n_1^2}-{n_2^2})-
({\vec n}_1-{\vec n}_2)({\vec\nu} \cdot{\vec n})\right]\cos(2\omega t)
\nonumber\\
&-6\xi_1\xi_2[{\vec n}\times({\vec n}_1 - {\vec n}_2)]
\left[ {\vec \nu} \cdot \left( {\vec e}_1 \times {\vec e}_2 \right) \right] 
\sin(2\omega t)
\biggr\}\, .
\end{align} 
\end{subequations}
We emphasize that the behaviour of ${\vec A}_1({\vec r},t)$ corresponds 
to the existence of radiation in the system, in analogy to 
the corresponding time-dependent term in Eq.~(\ref{phir});
however, the range of radial arguments $1/m\ll r\ll 1/\omega$ is
inappropriate for a discussion of this phenomenon.
The radial distances relevant for radiation 
will be discussed below in Sec.~\ref{photoncurrent}.

Using the formula ${\vec H} = \vec\nabla \times \vec A$,
applied to the results in Eqs.~(\ref{A0}) and~(\ref{A1}),
it is then very easy to obtain the corresponding asymptotics of the 
magnetic field in the relevant range of distances,
\begin{subequations}
\begin{eqnarray}
\label{H0}
{\vec H}_0({\vec r})&=&
\frac{3\alpha(Ze)\omega^2}{70\pi\,m^4r^4}
(\xi_1^2-\xi_2^2)\biggl\{5 \, {\vec n} \, {n_1 n_2}
\left[ {\vec \nu} \cdot \left( {\vec e}_1 \times {\vec e}_2 \right) \right] 
[({\vec n}_1- {\vec n}_2)\times{\vec\nu}]\biggr\}\,,\\
\label{H1}
{\vec H}_1({\vec r},t)&=&
\frac{\alpha(Ze)\omega^2}{20\pi\,m^2r^2}
\biggl\{\frac{11}{9}(\xi_1^2-\xi_2^2)
\left[{\vec n}\times{\vec\nu}\right]\cos(2\omega t)
\nonumber \\
&&+
(\xi_1^2+\xi_2^2)\,{\vec n}\,{n_1 n_2} \,
\left[ {\vec \nu} \cdot \left( {\vec e}_1 \times {\vec e}_2 \right) \right] \,
\cos(2\omega t)
-\xi_1\xi_2 {\vec n}\,
\left[ {\vec \nu} \cdot \left( {\vec e}_1 \times {\vec e}_2 \right) \right] \,
({n_1^2}-{n_2^2})\sin(2\omega t)
\biggr\}\, .
\end{eqnarray} 
\end{subequations}
At small distances, $r\ll 1/m$, the leading term of the 
asymptotics of the vector potential and the magnetic
field is time-independent:
\begin{eqnarray} 
\label{H0close}
{\vec A}_0({\vec r})&=&
-\frac{\alpha(Ze)\omega^2}{360\pi\,m^2r}
\biggl\{11(\xi_1^2+\xi_2^2)
\left[{\vec \nu}+{\vec n}({\vec \nu} \cdot {\vec n})\right]+
3(\xi_1^2-\xi_2^2)\,[{\vec n}\times[{\vec \nu}\times({\vec n}_1-{\vec n}_2)]]
\biggr\}\, ,\\
{\vec H}_0({\vec r})&=&
\frac{\alpha(Ze)\omega^2}{180\pi\,m^2r^2}
\biggl\{11(\xi_1^2+\xi_2^2)\left[{\vec n}\times{\vec\nu}\right]+
9(\xi_1^2-\xi_2^2)\,{\vec n}\,{n_1 n_2} \,
\left[{\vec \nu} \cdot ({\vec e}_1\times{\vec e}_2) \right]\biggr\}\, .
\end{eqnarray} 
\end{widetext}
In passing, we note that ${\vec H}_0({\vec r})$ of course has the 
transformation properties of a pseudovector, which is consistent 
with the necessary introduction of the factor 
${\vec \nu} \cdot ({\vec e}_1\times{\vec e}_2)$ in 
Eq.~(\ref{H0close}).
Also note that at all distances, ${\vec H}_0({\vec r})$ is an odd 
function of the vector $\vec n$. Therefore, due to the parity 
conservation, for any hydrogenic atomic state,
the energy shift related to the magnetic field vanishes 
in first order of perturbation theory.

%
%
\subsection{Application: transition probability in atoms}
\label{inducedtrans}

As shown in previous sections of this article,
the laser-dressed vacuum polarization induces additional 
electric and magnetic fields which modify the 
atomic transition probabilities.
We perform the calculation to the 
first order of perturbation theory with respect to these fields. The most 
interesting situation occurs when the laser frequency $\omega$ is closed 
to $\Delta E/2$, where $\Delta E$ is the energy interval between levels,
because the second-harmonic generation from the vacuum 
in this case leads to the generation of resonance photons. 
For simplicity, we restrict ourselves by the case of {\em circular}
polarization and transition $1s\rightarrow 2p$ (which corresponds to the 
electric dipole transition for spontaneous radiation).  We direct the
quantization axis along the vector $\vec \nu$. Due to 
the conservation of angular momentum, the matrix 
element is nonzero if $\Delta j_z= \pm 2$ only. The sign in this 
selection rule depends on the helicity of the laser wave.

Taking into account the selection rule, let us consider, for instance, the 
case of a 
right helicity of the laser wave and the transition from the state 
$1s_{1/2}$ with $j_z=-1/2$ to the state $2p_{3/2}$ with $j_z=3/2$. 
Because the induced potential $\Phi_1({\vec r}, t)$ is an even function of the 
vector $\vec n$, see Eq.~(\ref{Phi}), this potential does not 
contribute to the matrix element to the first order of perturbation 
theory, and the leading nonzero contribution comes from the perturbation 
Hamiltonian 
\begin{equation}\label{Pert}
V= -\frac{e}{m}({\vec A}_1({\vec r}, t){\vec p})
-\mu ({\vec \sigma}{\vec H}_1({\vec r}, t))\, , 
\end{equation}
where $\mu = e/(2m) = -|e|/(2m)$ is the electron 
magnetic moment, and ${\vec p}$ is the 
momentum operator. For the vector potential and the magnetic field, we 
should use its asymptotics at $1/m\ll r\ll 1/\omega$, 
as given in Eqs.~(\ref{A1}) and 
(\ref{H1}). For $s$ states, we can use 
${\vec p}\,\psi_{1s}(r)= im(Z\alpha){\vec n}\,\psi_{1s}(r)$, 
and for circular polarization with $\xi_1^2 = \xi_2^2$,
we have ${\vec n} {\vec A}_1({\vec r}, t)=0$ in the asymptotics (\ref{A1}).
In this setting, only the second term in the 
interaction Hamiltonian in
Eq.~(\ref{Pert}) gives a nonzero contribution. For the 
latter, a straightforward calculation of the matrix element gives
\begin{eqnarray}
\label{DeltaM}
&& \int\limits_{-\infty}^\infty dt \left<2p_{3/2},j_z\!=\!3/2|
\left[ -\mu{\vec \sigma}{\vec H}_1 \right]|1s_{1/2}, j_z\!=\!-1/2\right>
\nonumber \\
&&=-\frac{\alpha(Z\alpha)^3\,\omega^2\,\xi^2}{675\pi m} \,
2\pi\delta(E_{2p}-E_{1s}-2\omega)\, ,
\end{eqnarray}  
where ${\vec H}_1 \equiv {\vec H}_1(\vec{r},t)$ and 
$\xi^2 = \xi_1^2 = \xi_2^2$ for circular polarization.
Note that we here assume the quantization axis for the 
total electron angular momentum $j$ to be aligned along the
laser propagation (not polarization) direction, and the angular momentum 
projection $j_z$ to be defined with respect to that axis.
The resonance frequency for a hydrogenic 
transition involving a change in the principal quantum number
is of order $(Z\,\alpha)^2$, so that the prefactor in 
Eq.~(\ref{DeltaM}) is in fact of order $\alpha\,(Z\,\alpha)^7 \,\xi^2$,
where in turn $\xi^2$ is proportional to the laser intensity
according to Eq.~(\ref{parameters}).

The transition $1s \to 2p_{3/2}$ is driven at the 
second harmonic of the incident laser frequency [see the 
Dirac delta function $\delta(E_{2p}-E_{1s}-2\omega)$], anticipating the 
possibility of photon fusion in a strong laser field, to be discussed
below in Sec.~\ref{photoncurrent} in the far field $r \gg 1/\omega$
from the atomic nucleus.

The transition amplitude in Eq.~(\ref{DeltaM}) should now be compared to other
effects which also can induce small nonvanishing  
transition probabilities for $1s_{1/2} \to 2p_{3/2}$,
even if the incident radiation is off resonance and
amounts to only half the frequency of the resonance transition.
One such term is given by
\begin{widetext}
\begin{equation}
\label{nonvanishing}
\left< 2p_{3/2},j_z\!=\!3/2 \left| \frac{e}{m}({\vec A}({\vec r}, t){\vec p})\,
\frac{1}{E - H - \omega}\, 
\mu ({\vec \sigma}{\vec H}({\vec r}, t)) \right| 1s_{1/2}, j_z\!=\!-1/2 \right>
\end{equation}
\end{widetext}
Note that the fields ${\vec A}({\vec r}, t)$ and ${\vec H}({\vec r}, t)$
are the full laser fields, i.e.~they do not carry any subscripts 
like the vacuum-polarization induced fields. 
For a transition involving angular momentum projections
$j_z=1/2 \to j_z=3/2$ (in contrast to 
$j_z=-1/2 \to j_z=3/2$), the term corresponding to (\ref{nonvanishing})
with two interactions $-\frac{e}{m}({\vec A}({\vec r}, t){\vec p})$ 
may contribute, when we additionally
consider higher-order multipoles of the laser field.
Note that the laser field uniquely specifies a 
polarization axis along which the exponential $\exp(i \vec{k}\vec{r}) =
\exp(i \omega \vec{\nu}\vec{r})$ should be expanded.

Using power counting and the $Z\,\alpha$-expansion, it is easy to show 
that the matrix element in Eq.~(\ref{nonvanishing}) is of order
$Z\alpha\,\xi^2$, which leads to a relative factor of 
\begin{equation}
\label{r}
r = \frac{\alpha\,(Z\,\alpha)^7}{Z\,\alpha} = \alpha\,(Z\,\alpha)^6 
\approx 10^{-15} \, Z^6\,.
\end{equation}
for the ratio of the vacuum-induced to the second-order induced 
transition probability.
The ratio $r$ is very small for $Z = 1$, and it is still 
only of order $10^{-8}$ for a medium-$Z$ range ($Z \approx 20$).
While it is not negligibly small for heavy ions,
the transition frequencies increase with increasing $Z$,
and powerful coherent lasers in this range of frequencies are
still unavailable. 

In order to observe the phenomenon of 
influence of second-order harmonic generation 
due to laser-dressed vacuum polarization on the transition probability, 
it could be useful explore whether transitions exist where the 
ratio $r$ is more favourable with regard to a possible experimental
investigation. In this case, a very careful estimate of the background would 
be required. For instance, the linewidth of even narrow 
lasers can induce very small, but important transition 
probabilities far off resonance, and these would also 
have to be taken into account in such an analysis.

%
%
\subsection{Application: photon fusion}
\label{photoncurrent}

Due to the vacuum polarization in the Coulomb field, nonlinear QED 
processes such as Delbr\"uck scattering (coherent photon scattering in the 
atomic electric field) and photon splitting (transition of initial photon 
into two final photons) appear. These processes are investigated both 
theoretically and experimentally, see 
Refs.~\cite{PaMo1975,MiSc1994,AzEtAl2002,LeEtAl2003}.
Another closely related process is the photon fusion in an atomic 
field. In this process, two initial photons with the energy $\omega_1$ and 
$\omega_2$ transfer to the one final photon with the energy 
$\omega_3=\omega_1+\omega_2$. This process has not been observed 
experimentally because of its small probability. 

The existence of the radiation phenomenon has been suggested by the 
asymptotic structure of the time-dependent components of the 
induced scalar potential in Eq.~(\ref{phir}) and the induced vector potential 
in Eq.~(\ref{A1}). Photon fusion takes place when two laser photons 
with the energy $\omega$ become a single photon of frequency
$2\omega$, and this process is otherwise known as second-order 
harmonic generation from the vacuum in the presence of a 
Coulomb field. 

The probability of this process can be calculated as follows.
First, it is necessary to calculate the electromagnetic field at 
distances $r\gg 1/\omega$, characteristic of radiation. 
A straightforward calculation leads to the 
following asymptotics for the potential $\Phi({\vec r},t)$ and the vector 
potential ${\vec A}({\vec r},t)$:
\begin{widetext}
\begin{align}
\Phi({\vec r},t)=&
\frac{\alpha(Ze)\omega^2}{720\pi\,m^2}
\biggl\{3(\xi_1^2+\xi_2^2)({n_1^2}-{n_2^2})
-11(\xi_1^2-\xi_2^2)({n_1^2}+{n_2^2})
+12\,i\,\xi_1\xi_2{n_1 n_2}\biggr\}\,
\frac{\exp[2i\omega(r-t)]}{r(1-({\vec \nu} \cdot {\vec n}))}\,\,+
\,\,\mbox{C.C.}\,,\nonumber\\
{\vec A}({\vec r},t)=&
\frac{\alpha(Ze)\omega^2}{720\pi\,m^2}
\biggl\{3(\xi_1^2+\xi_2^2)\left[{\vec\nu}\,({n_1^2}-{n_2^2})
+({\vec n}_1-{\vec n}_2) \,
(1- ({\vec \nu} \cdot {\vec n}))\right]\nonumber\\
&-11(\xi_1^2-\xi_2^2)
\left[{\vec\nu}\,({n_1^2}+{n_2^2}) +
({\vec n}_1+{\vec n}_2)(1- ({\vec \nu} \cdot {\vec n}))\right]
\nonumber\\
&+6\,i\,\xi_1\xi_2\,
\left[({\vec \nu}-{\vec n})\times
({\vec n}_1- {\vec n}_2)\right]
\left[ {\vec \nu} \cdot \left( {\vec e}_1 \times {\vec e}_2 \right) \right] \,
\biggr\}\,
\frac{\exp[2i\omega(r-t)]}{r(1-({\vec \nu} \cdot {\vec n}))}\,\,+
\,\,\mbox{C.C.}\,,
\end{align} 
where by C.C. we denote the complex conjugate.
Here, we omitted terms proportional to 
$\exp[\pm 2i(\,\omega \,t - {\vec\varkappa} {\vec r})]$.
These terms lead to a energy flux along the propagation 
direction of the laser field and ensure the fulfillment of the 
energy conservation condition.

From these formulas we obtain the corresponding expression 
for the magnetic field ($r \gg 1/\omega$),
\begin{eqnarray}
{\vec H}({\vec r},t)&=&
\frac{i\,\alpha(Ze)\omega^3}{360\pi\,m^2}
\biggl\{3(\xi_1^2+\xi_2^2)\left[2({\vec n}-{\vec\nu})\,
[{\vec\nu} \cdot ({\vec n}_1\times{\vec n}_2)]+
(1- ({\vec \nu} \cdot {\vec n}))
[({\vec n}_1-{\vec n}_2)\times{\vec \nu}]\right]\nonumber\\
&&-11(\xi_1^2-\xi_2^2)(1-({\vec \nu} \cdot {\vec n}))[{\vec n}\times{\vec \nu}]
\nonumber\\
&&+6\,i\,\xi_1\xi_2\left[({\vec \nu}-{\vec n})( n_1^2-n_2^2)+
({\vec n}_1-{\vec n}_2)(1-({\vec \nu} \cdot{\vec n})) \right] \,
\left[ {\vec \nu} \cdot ({\vec e}_1\times{\vec e}_2) \right]\,
\biggr\}\,
\frac{\exp[2i\omega(r-t)]}{r(1-({\vec \nu}\cdot{\vec n}))}\,\,+
\,\,\mbox{C.C.}\,  ;
\end{eqnarray} 
The magnetic field is perpendicular to the vector $\vec n$,
as it should be. For the electric field we obtain
\begin{equation}
{\vec E}({\vec r},t)=-{\vec \nabla}\Phi({\vec r},t)-
\frac{\partial}{\partial t}{\vec A}({\vec r},t)=
[{\vec H}({\vec r},t)\times {\vec n}]\, ,
\end{equation}
as should be for the field of radiation. Now we can calculate the 
differential distribution of the energy flux
\begin{eqnarray}\label{dS}
dS&=&\frac{{\vec n} \cdot 
\left[{\vec E}({\vec r},t)\times {\vec H}({\vec r},t)\right]}
{4\pi}\,r^2\,d\Omega
\nonumber\\
&&=\left(\frac{\alpha(Ze)\omega^3}{120\pi\,m^2}\right)^2
\biggl\{\left[(\xi_1^2+\xi_2^2)^2
+\frac{56}{9}(\xi_1^2-\xi_2^2)^2\right] (n_1^2+n_2^2)
-\frac{11}{3}(\xi_1^4-\xi_2^4)( n_1^2-n_2^2)\biggr\}\frac{d\Omega}{4\pi}\, .
\end{eqnarray} 
For the total energy flux we have
\begin{eqnarray}
S&=&\frac{2}{3}\left(\frac{\alpha(Ze)\omega^3}{120\pi\,m^2}\right)^2
\left[(\xi_1^2+\xi_2^2)^2+\frac{56}{9}(\xi_1^2-\xi_2^2)^2\right]
=\frac{\alpha(Z\alpha)^2\omega^2}{5400\pi^2}
\left(\frac{{\cal E}}{{\cal E}_0}\right)^4
\left[1+\frac{56}{9}\left(
\frac{\xi_1^2-\xi_2^2}{\xi_1^2+\xi_2^2}\right)^2\right] \, .
\end{eqnarray} 
\end{widetext}
In view of the suppression factor $({\cal E}/{\cal E}_0)^4$,
this flux is very small, and the observation of this phenomenon 
demands very intense laser sources. 
Note that the result (\ref{dS}) can be also 
obtained with the help of the Heisenberg-Euler effective Lagrangian (see, 
e.g., Ref.~\cite{BeLiPi1982}). We have checked by explicit 
calculation that our result 
(\ref{dS}) agrees with the alternative derivation
using the Heisenberg-Euler effective Lagrangian. 

%
%
\section{CONCLUSION}
\label{conclusion}

We have investigated, in detail, the properties of the (one-loop)
vacuum-polarization tensor in the presence of two nontrivial
background fields: (i) the atomic, nuclear Coulomb field, and (ii) 
a possibly intense laser field. The potential induced by the 
laser-dressed vacuum polarization in a Coulomb field
is derived in Sec.~\ref{density}.
The approximations employed in the calculation concern the frequency of the 
driving laser $\omega$, which we assume to be small as compared to 
$m\,c^2/\hbar$, where $m$ is the mass of the electron 
[see Eq.~(\ref{assump1})].
This condition is fulfilled by optical wavelengths and 
for x-rays. Another assumption concerns the 
laser field strength ${\cal E}$: our calculation is
valid provided that ${\cal E}/{\cal E}_0 \ll 1$ [see Eq.~(\ref{assump2})],
where ${\cal E}_0$ is the critical Schwinger electric field,
but they remain approximately applicable 
to situations where, say, ${\cal E}/{\cal E}_0 \approx 1/10$.
Consequently, our calculations remain valid for very intense 
laser fields, but they have to be modified 
when the amplitude of the oscillatory electric field is 
equal or even exceeds the Schwinger critical electric field.

After a derivation of the induced charge density in 
Sec.~\ref{calcdensity},
the asymptotics of the laser-induced potential are analyzed in 
Sec.~\ref{asymppotential}, both in a region close to the nucleus 
($r \ll 1/m$) and in a region which includes the 
Bohr radius ($1/m \ll r \ll 1/\omega$). The additional laser-induced 
potential induces energy shifts to atomic levels which 
are analyzed in Sec.~\ref{energypotential}.
We find that in addition to a familiar
$S$-state energy shift [see Eq.~(\ref{es})], 
which is present even without a laser field,
there is an additional term which is relevant for $P$ states and 
states with higher angular momenta,
which is generated by a quadrupole-type contribution
to the laser-induced vacuum-polarization potential
[see Eq.~(\ref{ep})]. 
The latter contribution generates a ``polarized Lamb shift''
which is akin to similar effects predicted in a noncommutative 
space-time~\cite{ChSJTu2001}. 

The analysis of the charge density and the potential induced 
by the laser-dressed vacuum-polarization is only a part of the 
solution of the problem (see Sec.~\ref{current}). 
The laser field breaks the rotational symmetry of the 
pure Coulomb potential and induces, in addition, a nonvanishing current 
density and a vector potential (see Sec.~\ref{calccurrent}), 
which gives rise to additional fields, which have both time-independent and 
oscillatory components. The time-dependent electric and magnetic components
oscillate at the second harmonic of the incident laser
frequency and are analyzed in a region which includes 
the Bohr radius of a hydrogenlike system (see Sec.~\ref{asympvectorpot}). 
The time-dependent 
electric and magnetic components can lead to electric-dipole transitions
via second-order harmonic generation from the vacuum
(see Sec.~\ref{inducedtrans}).
In the far-field, the induced electric and 
magnetic fields correspond to the radiation of 
real photons, at the second harmonic of the laser
(see Sec.~\ref{photoncurrent}). This process is suppressed
by a factor $({\cal E}/{\cal E}_0)^4$ and therefore
requires excessively large laser field strengths to be 
experimentally significant.

Two points are worth mentioning here: (i) all effects derived in
the current investigation are suppressed at least by the 
square of the parameter $({\cal E}/{\cal E}_0)^2$.
This suppression is directly related to the fact that the 
virtual particles in the vacuum-polarization loop are
effectively ``detached'' from the atomic physics energy scale
on which the dominant laser-dressed radiative effects are 
commonly expected~\cite{JeEvHaKe2003,JeKe2004aop,EvJeKe2004,JeEvKe2005}.
This suppression leads to the conclusion
that the effects will be difficult to observe, experimentally,
even at the current high-intensity laser facilities.
However, with the advent of ever more powerful laser facilities 
which can generate intense pulses, the effects might become 
within the reach of experimentalists in the not too distant future.
On the basis of the Heisenberg--Euler Lagrangian, the suppression 
could have been expected, since the additional terms which 
supplement the Maxwell Lagrangian are also suppressed by
this factor. The Heisenberg--Euler Lagrangian
is intimately related to the problem studied here 
(this has been pointed out in Sec.~\ref{photoncurrent}).

One point in particular seems worth a note:
The ``polarized Lamb shift''~\cite{ChSJTu2001} which is 
expected in a noncommutative space-time, finds a natural
counterpart in terms of a laser-induced effect which is always
present when high-precision spectroscopy experiments are 
performed [see Eq.~(\ref{ep})]. 
This observation puts a severe restraint on the observability
of noncommutative  space-time effects in atomic spectroscopy.

In our calculation, we encounter additional finite renormalizations
which have to be carried out although the vacuum-polarization
tensor, which forms the starting point of our calculation 
is in principle already renormalized on mass shell 
[see Eqs.~(\ref{subtractrho}) and (\ref{subtractJ})]. The reason 
is that the additional external Coulomb field induces spurious 
effects, which have to be eliminated based on the conservation 
conditions for the electric charge and the nonexistence of a
magnetic monopole. The necessity of carrying out additional
finite subtractions is akin to the additional subtractions that have to be 
carried out in the calculation of the Wichmann--Kroll 
higher-order one-loop vacuum-polarization corrections in atoms,
where, indeed, additional subtractions have to be carried out in each 
order of the $Z\,\alpha$ expansion.

Our investigation has been carried out in the more 
general context of quantum field theory under the influence 
of external (``unusual'') conditions. We have analyzed the 
vacuum polarization tensor in the presence of a laser field 
and a Coulomb field, which mediate quantum electrodynamic 
effects under the influence of dynamically changing external 
conditions. A few curious phenomena are encountered: 
finite renormalizations that apply to the induced charge and 
current density [see Eqs.~(\ref{subtractrho}) and~(\ref{subtractJ})], 
induced charge distributions with an unusual
asymptotic behaviour [see Eqs.~(\ref{phir}), (\ref{A0}) and~(\ref{A1})], 
and the rotational symmetry breaking of the quantum electrodynamic 
effects [see Eq.~(\ref{ep})]. The techniques outlined here might be useful in 
other situations where our understanding is more limited.
In particular, we have shown that general wisdom regarding the asymptotic
structures of the quantum electrodynamic effects has to be modified 
quite drastically in the presence of further external conditions,
like a laser field.

%
%
\section*{ACKNOWLEDGMENTS}

A.I.M. and I.S.T. thank the Max-Planck-Institute for Nuclear Physics, 
Heidelberg, for warm hospitality on the occasion of an extended guest 
researcher appointment during which this work was performed.
Partial support by RFBR Grants 03-02-16510 and 
05-02-16079 is also gratefully acknowledged. 
The authors acknowledge helpful conversations with 
Professor G. Kryuchkyan at an early stage of this
project, and helpful discussions M. Haas. Additionally,
the authors would like to thank M. Haas for providing Figs.~\ref{isosurf}
and~\ref{projections}.

\end{document}